*Article*

# Automatic Staff Reconstruction within SIMSSA Project

**Lorenzo J. Tardón** [1,*] , **Isabel Barbancho** [1,*] , **Ana M. Barbancho** [1] **and Ichiro Fujinaga** [2]

[1] ATIC Research Group, Dpto. Ingeniería de Comunicaciones, Universidad de Málaga, Andalucía Tech, Campus Universitario de Teatinos s/n, 29071 Málaga, Spain; abp@ic.uma.es (A.M.B.); ichiro.fujinaga@mcgill.ca (I.F.)

[2] Schulich School of Music, McGill University, Montreal, QC H3A 1E3, Canada

* Correspondence: lorenzo@ic.uma.es (L.J.T.); ibp@ic.uma.es (I.B.); Tel.: +34-952-13-11-88 (L.J.T.); +34-952-13-25-87 (I.B.)

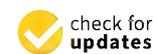

**Abstract:** The automatic analysis of scores has been a research topic of interest for the last few decades and still is since music databases that include musical scores are currently being created to make musical content available to the public, including scores of ancient music. For the correct analysis of music elements and their interpretation, the identification of staff lines is of key importance. In this paper, a scheme to post-process the output of a previous musical object identification system is described. This system allows the reconstruction by means of detection, tracking and interpolation of the staff lines of ancient scores from the digital Salzinnes Database. The scheme developed shows a remarkable performance on the specific task it was created for.

## 1. Introduction

Music is a main source of the world's cultural heritage that, since ancient times, has been transmitted and stored by handwritten and printed music scores thanks to the inherent bond between the sound of music and its written representation. This is why the automatic processing, analysis and interpretation of music scores, Optical Music Recognition (OMR), has been a research topic of interest for decades [1–3] and it is still of great relevance thanks to the desire to preserve our cultural heritage and, more specifically, our musical heritage by means of the creation of digital databases with musical content, including digitized scores.

In this context, the need for transcription of musical scores to aid their distribution, understanding and dissemination is clear. Unfortunately, manual transcription is a very much time-consuming task. This fact leads to the interest on the development of OMR algorithms and related tools to facilitate search, analysis and interpretation of the scores.

Commonly, OMR systems include a series of stages for image pre-processing [4,5], staff detection and removal [6,7], musical symbol isolation and recognition [8,9] and musical information reconstruction and representation [3,10].

Music information reconstruction includes the interpretation of the meaning of extracted musical symbols and, specifically, their relative position with respect to the corresponding staff. The detection and classification of musical symbols has been faced by using different approaches [11,12] with neural-network-based approaches coming out recently [8,13,14], which often require large datasets for training [15]. However, only recently some of these works specifically consider the full extraction of staff lines [16] or the precise identification of their location [17].

For the correct interpretation of musical symbols, their location with respect to the staff is of main importance and, consequently, the staff lines need to be detected as accurately and completely as possible. This is a task that is not often considered in OMR systems. Conversely, the actual objective related to the processing of staff lines is commonly their removal in the first stages of the processing schemes to facilitate the later identification of other musical objects [6,18] since they are often considered an obstacle for symbol segmentation recognition [19].

In this paper, we propose a processing scheme that performs the reconstruction of the staves in processed images of actual ancient scores containing partially deleted staves, the remaining or portions of symbols from a previous symbol extraction scheme and other artifacts.

The paper is organized as follows. In the next section, the context that defines the type of images we are dealing with is established. In Section 3, the complete scheme for staff detection and tracking aimed at staff reconstruction is presented. Section 4 contains an analysis of the behavior of the proposed scheme, including results and a discussion. Finally, conclusions are drawn in Section 5.

## 2. Context: SIMSSA Project

Within the context of cultural preservation and dissemination, the Single Interface for Music Score Searching and Analysis (SIMSSA) project [20] aims at not only making the digital images of ancient scores available to the public but, also, analyzing their structure to make them accessible. Thus, the interpretation of the meaning of musical symbols is of great importance and so is the correct identification of their location with respect to their corresponding staff. Among others, the project deals with the music scores in the Salzinnes Antiphonal manuscript dated to the middle of the 16th century; this is the data that will be dealt with within this work. Different schemes have been used to process and analyze these data: convolutional neural networks have been employed for binarization [5], categorization of score elements [14,21] and layering into their constituent elements [22]. An example of the application of this analysis process is shown in Figure 1, where the original image (a), staves (b), background (c) and symbols extracted (d) are shown for a sample score. In these approaches, the staff lines are still considered to make the detection, segmentation, and classification of symbols more complicated, which encourages their removal.

The detection and complete tracking and interpolation of the staff lines remains unconsidered in spite of its importance for the interpretation of the symbols [14]. The reconstruction of staff lines is the main focus in this work.

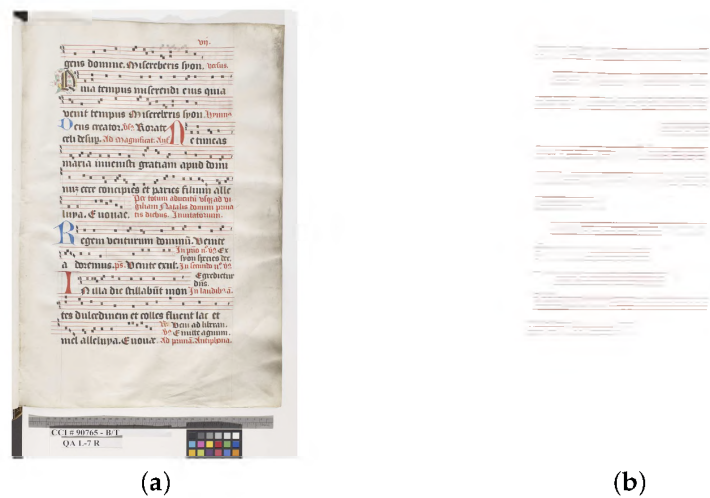

(a)  (b)

**Figure 1.** *Cont.*

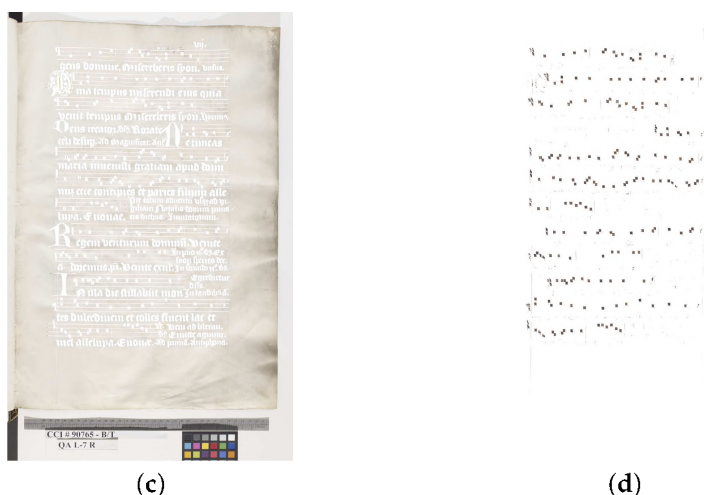

(c)                                        (d)

**Figure 1.** Example of application of the Pixelwise scheme on folio 7r of Salzinnes book. (**a**) Scanned image. (**b**) Extracted staves. (**c**) Image background. (**d**) Musical symbols extracted.

## 3. Staff Detection and Tracking System

The staff tracking algorithm is made up of several subsystems aimed at preparing the input image for performing the actual tracking and adjust the generated path to a smooth curve. The parameters are established taking into account their relative value with respect to the score image size and the specific task at hand. Thus, the parameters are chosen specifically for the book under consideration (see http://salzinnes.simssa.ca/). Concretely, the size of score pages is approximately 61.5 cm × 39.5 cm and they were scanned with a resolution of $N \times M = 6993 \times 4414$ pixels; the scores contain four-staff-line music notation with a separation between consecutive staff lines around 40–60 pixels.

When the staff image (Figure 1b) obtained by the Pixelwise scheme is fed into the tracking scheme, a number of stages will be performed that can be grouped into the following main tasks:

- Pre-processing.
- Search for staves.
- Track staff lines.
- Fit a smooth model to staff lines.

These main stages contain several sub-tasks that will be described next. Note that the search-for-staves stage is an iterative process performed until no more staves are found. The detection and complete tracking of a certain staff implies its removal from the image to allow the next staff search process. This procedure, together with the main scheme stages, is shown in the diagram depicted in Figure 2.

The system stages, their sub-tasks and their parameters will be described in the next sections. Note that the system parameters are mainly based on the relation between the scanning resolution and the size of the music objects and, specifically, on the observed area of the smallest object detected previously, the expected range of the staff line separation and its estimation by the scheme itself.

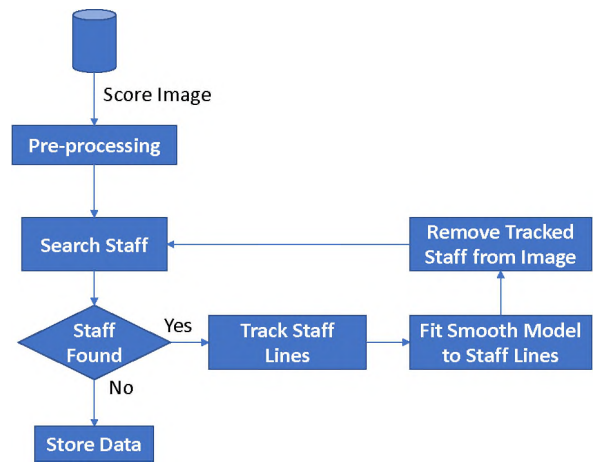

**Figure 2.** General scheme of the staff reconstruction system.

## 3.1. Pre-Processing for Staff Tracking

The first stage guarantees that the input image for the rest of the process is a binary one with black background so that the morphological operators and counting stages perform as expected. Thus, the output images from the Pixelwise scheme, $I(n, m)$, containing pieces of the staff lines are binarized by making use of the Otsu's method [23]:

$$I_b(n, m) = O_b(I(n, m)) \tag{1}$$

where $O_b(\cdot)$ stands for Otsu's binarization scheme. Note that this is not a main stage because the data have been previously processed [5]. Nevertheless, since we need binary images as input, such a process needs to be carried out.

The rest of the pre-processing scheme is aimed at removing small unconnected detections. To this end, the objects in the image were observed. Ideally, the Pixelwise algorithm removes all elements in the score and should return a set of unconnected broken staff segments, however, numerous other artifacts remain. Nevertheless, we expect each staff line piece to have the structure of a line segment of a minimum size built upon pixels connected in vertical, horizontal or diagonal (8-connected). This minimum size was defined as an area of $\lambda = 500$ pixels; this is approximately 20% of the area of the smallest element removed by the previous scheme, at the working resolution. This parameter is employed to apply an area opening algorithm [24], $\gamma(\cdot)$, on the image:

$$I_o(n, m) = \gamma(I_b(n, m), \lambda) \tag{2}$$

to remove small artifacts. Thus, $I_o(n, m)$ contains all the 8-connected components in $I_b(n, m)$ with area larger than $\lambda$.

Then, a morphological closing step [25] with a disc-shaped structuring element of the same area as in the area opening stage, $\lambda$, is performed [26]:

$$I_p(n, m) = I_o(n, m) \bullet D \tag{3}$$

where $\bullet$ denotes the morphological closing operation and $D$ the disc-shaped structuring element of area $\lambda$. Figure 3 shows an illustration of the result of the application of these steps. The resulting image contains much fewer artifacts than the original one, although some small staff line segments can be missed.

Note that, actually, there is no strict need for the area opening and morphological closing steps, however, their usage helps to obtain better estimations of the staff lines since they are less likely distorted due to segments of letters or drawings not removed by the object detection algorithm. Figure 4 illustrates the influence of spurious pieces of objects left by the previous algorithm on the staff tracking scheme.

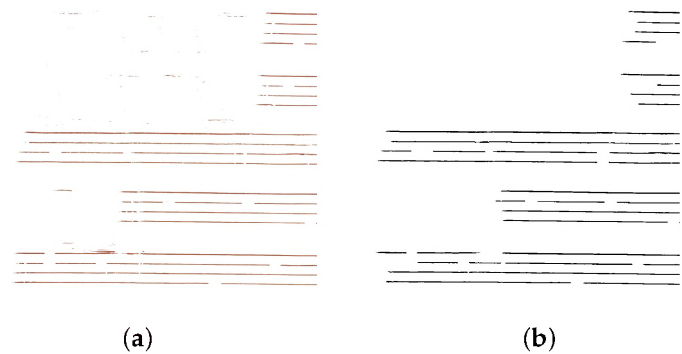

| (a) | (b) |

**Figure 3.** Image detail of folio 1r of Salzinnes book illustrating the application of the binarization and cleaning stages. (**a**) Input image. (**b**) Image after the pre-processing stage.

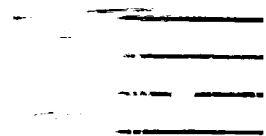

(**a**) Spurious marks left by the object detection algorithm.

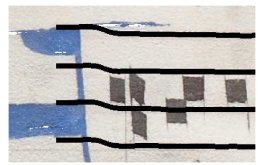 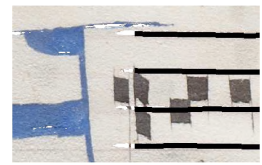

(**b**) Result of the staff tracking algorithm when the opening and closing steps are not performed.

(**c**) Result of the staff tracking algorithm when the opening and closing steps are carried out.

**Figure 4.** Illustration of different results when opening and closing pre-processing steps are not applied or applied before the staff tracking algorithm (described in Section 3.3). Detail of folio 52r of Salzinnes book.

### 3.2. Staff Search

After the image has been pre-processed, a procedure to identify image excerpts that contain proper tetragrams is applied. Note that we are dealing with a specific type of scores in which the music is coded in tetragrams instead of the pentagrams commonly used in modern Western music.

The staff search scheme is based on the utilization of the horizontal histogram or y-projection [27,28] of the clean image. The image is first divided into a number of vertical stripes: $N_s = 16$ stripes, though this number should become larger to detected small pieces of staff lines that by no means fill the page horizontally. Nevertheless, we found that the selected number performs well in most of the observed cases under consideration. The horizontal histogram counting the number of white pixels, $Y_h(n)$, is obtained recursively for each stripe, until a staff is found:

$$Y_h(n) = \sum_{j=1}^{M} I_p(n,j) \qquad (4)$$

The histogram is smoothed by using a moving average filter, $h(n)$, of odd length not smaller than the approximate thickness of the staff lines; in our case, the length chosen is $l = 11$, which is approximately the observed mean thickness of a staff lines. Then:

$$Y_s(n) = Y_h(n) * h(n) \tag{5}$$

Afterward, all the bins in the histogram with too few samples to correspond to a staff line are set to zero:

$$F_s(n) = \begin{cases} Y_s(n), & Y_s(n) > h_{th} \\ 0, & \text{otherwise} \end{cases} \tag{6}$$

The threshold is manually set to 40% the stripe width: $h_{th} = 0.4 \frac{N}{N_s}$. Note that it is very highly improbable that horizontal lines not containing a staff line (or part of it) may exceed this threshold because of the nature of the scores themselves (see Figure 1). Then, local maxima are extracted from $F_s(n)$. The maxima must be separated at least the minimum number of pixels of separation between the staff lines to ignore less important peaks in the vicinity of a larger local peak.

Later, the scheme searches for four consecutive local maxima with the expected structure of a staff, i.e., four lines separated a similar distance within a certain range. In our case, the range selected is [20–100] pixels at the current resolution for the book under analysis. This range widely contains the expected range of staff line separation values in the scores to analyze. Then, if the separation between the selected maxima is similar, i.e., the separation between each pair of hypothetical consecutive staff lines, $\Delta_i$, with $i = 1, 2, 3$, does not differ more than 20% from the average separation between the other tentative staff lines in the assumed staff under test,

$$\overline{\Delta_i C} = \frac{1}{2} \sum_{\substack{j=1,\dots3 \\ j \neq i}} \Delta_j, \tag{7}$$

then a staff is detected at the present stripe. This difference would be easily visually observable and, thus, it should not exist in the drawn staves. This information is used to immediately start a two-way staff line tracking and to estimate the separation between the current staff lines:

$$\overline{\Delta} = \frac{1}{3} \sum_{j=1}^{3} \Delta_i \tag{8}$$

If no staff is detected in a certain stripe $s$, another one will be tested until all of them are evaluated; if no staff is found in any stripe then the staff search algorithm ends. The overall procedure is depicted in Figure 5.

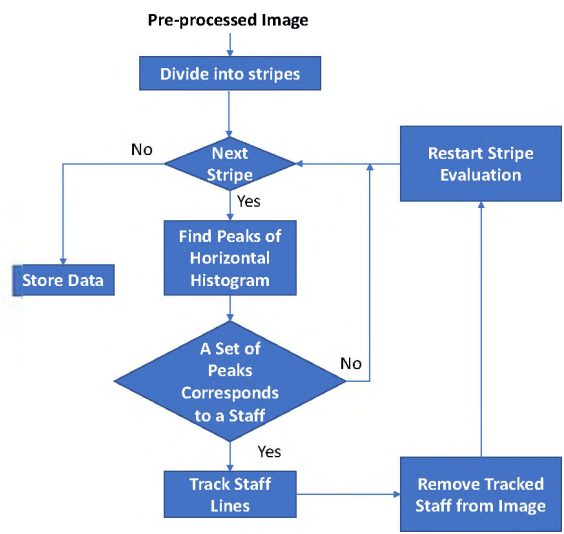

**Figure 5.** Diagram of the staff search scheme.

*3.3. Staff Line Tracking*

The staff line tracking algorithm will be performed both from left to right and right to left directions. Later, the best one of the resulting tracked lines will be initially chosen and later improved by the data from the other outcome of the tracking scheme. This selection will be taken in terms of a measurement of the estimated validity of the specific tracking; this idea will be described later.

Considering left-to-right tracking, the initial search point is obtained find finding a column such that four segments, one for each of the staff lines, are found filling a certain tile. The starting point will be situated at the vertical center of such tile. We have selected this tile to be $\frac{2}{3}\overline{\Delta}$ long and 3 px high in order to ensure that the starting tracking points effectively lie on the staff lines but not on their edge (note that height chosen is the minimum value which guarantees that the starting point will be on the staff line but not on its upper or lower edge).

Then, the vertical center of each staff line $i$, $g_i(m)$, will be tracked column by column toward the right. If a certain staff line is actually detected, the vertical position of the center of the staff is found, stored as new tracked staff line position and marked as valid detection. Otherwise, the vertical position of the staff line is neither modified nor marked as valid detection:

$$v_i(m) = \begin{cases} 1, & \text{valid detection} \\ 0, & \text{otherwise} \end{cases} \tag{9}$$

Then, each staff line that could not be detected is checked and the update of their vertical position, $g_i(m)$, is done taking into account the remaining detected staff lines. Specifically, the mean update of the vertical position of the successfully detected staff lines,

$$\delta g_i(m) = g_i(m) - g_i(m-1), \tag{10}$$

with $v_i(m) = 1$, is employed to update the vertical position of the undetected ones (those with $k|v_k(m) = 0$) as follows:

$$\delta g_k(m) = \frac{1}{N_{v_i(m)}} \sum_{i|v_i(m)=1} \delta g_i(m) \tag{11}$$

Their position is marked as estimated by using other staff lines; the number of staff lines employed to make the estimation is also stored.

Note that the location of the staff lines is tracked using floating point precision and the pixels that define the staff lines are required to be 8-connected.

After all the staff lines at a position are considered, it is checked whether their update was done supported by at least one detection or not. If not, then, the update of their vertical position is calculated by using the mean slope of the tracked staff lines of the current stave using previous detected positions. Specifically, the previous $S$ pixels, with $S$ half the estimated separation between the staff lines are used, let $\mathcal{S}$ denote this set of pixels; then, for a certain staff line,

$$\delta g_i(m) = \frac{\sum_{r \in \mathcal{S}} \delta g_i(r)}{S} \tag{12}$$

If no sufficient data are available, missing data are considered to define a horizontal line.

Then, all the staff lines are updated using:

$$\delta \overline{g}(m) = mean(g_i(m)) \tag{13}$$

Additionally, if no detection for the current column is found, the stored update of the vertical position of the staff lines $\delta g_i(m)$ for subsequent calculations of the slope is set to zero, which adds a tendency toward horizontal staff lines, in accordance with the fact the score images had their rotation originally corrected.

A simplified diagram of the tracking process is shown in Figure 6; only the main stages and decisions are shown.

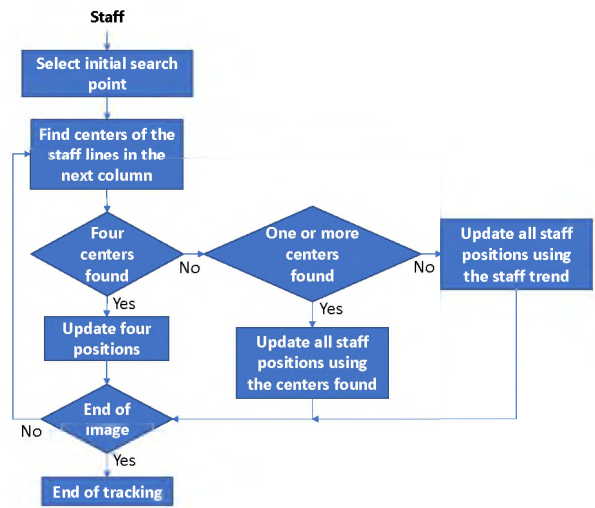

**Figure 6.** Simplified diagram of the staff tracking scheme.

After the procedure described runs in the two horizontal directions, the outcome with more valid positions is selected as principal. The staff line pixels in the principal outcome that were either not detected or tracked with only one support staff line are updated making use of staff line pixels of the other outcome when the latter come from actual detection or estimation based on at least two other staff line pixels.

A horizontal image stripe containing the detected staff is deleted from the image to allow the recursive detection of remaining staves.

### 3.4. Staff Line Smoothing

After the tracking process is complete for a selected staff, its resemblance can be jagged at certain points due to false detections, errors, etc., but also to the task itself, i.e., an approximately horizontal segment could be tracked by a saw-tooth pattern due to nature of the discretized images themselves (see Figure 7c). This type of pattern is not expected for a staff line. To avoid this, a smoothing spline is constructed for each of the staff lines found. The smoothed staff line $i$ is the curve $s_i^*(m)$ that minimizes [29]:

$$s_i^*(m) = \min_{s_i(m)} \left( p \sum_m (g_i(m) - s_i(m))^2 + (1 - p) \int_{\mathcal{C}_m} \left( \frac{d^2 s_i(x)}{dx^2} \right) dx \right) \tag{14}$$

where the smoothing parameter is $p = 10^{-4}$. The second derivative is obtained at the data points by means of its discrete approximation [29]. After the procedure is complete, the curve data are rounded to the nearest integer:

$$L_i(m) = round(s_i^*(m)) \tag{15}$$

Note that this procedure will also help remove some isolated spurious detections.

Figure 7 illustrates the application of the smoothing scheme. Figure 7a displays a detail of an original image composed of four staff lines. Figure 7b shows an enlarged view of the fourth staff line displayed in Figure 7a. Figure 7c presents the corresponding tracked staff line, this image shows clearly the saw-like structure of a two pixel wide tracked line derived from the uncertainty of the location of the center of the staff line due to its scanned thickness. Figure 7d shows the final result for the same staff line after the smoothing operation: the staff line finally obtained is a much better representation of the original staff line.

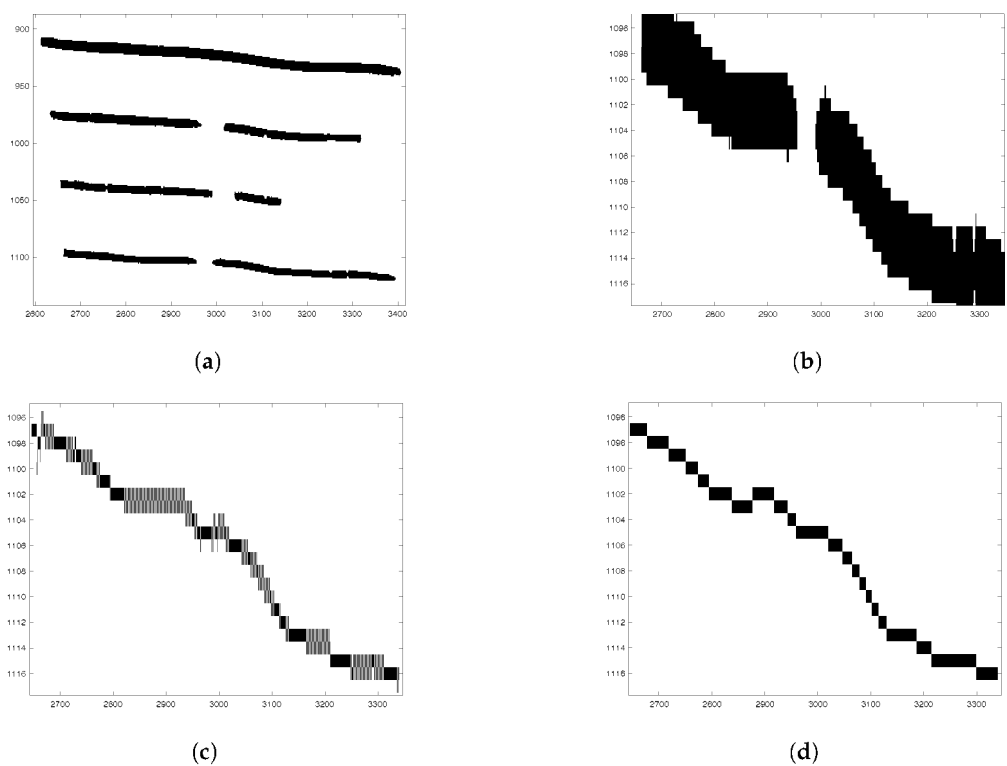

**Figure 7.** Illustration of the performance of the smoothing procedure. (**a**) Detail of the first staff of the original binary image corresponding to folio 2r. (**b**) Detail of the lowest staff line of (**a**). (**c**) Detail of the tracked and interpolated staff line for the image in (**b**). (**d**) Detail of the smooth reconstructed staff line for the image in (**b**).

### 3.5. Staff Image Creation

After all the tetragrams have been found, or no more are detected by the staff search scheme, an image that represents the reconstructed staff lines is created for each page analyzed.

Figure 8 illustrates the result of drawing the reconstructed staff lines with the estimated line thickness over the original image.

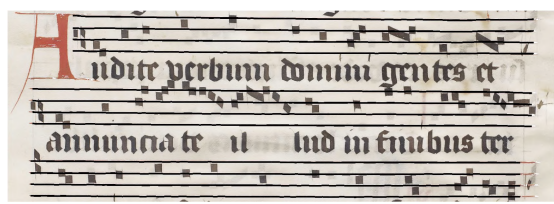

**Figure 8.** Reconstructed staff lines drawn over the original image analyzed. Detail of folio 4r.

## 4. Results and Discussion

The method described has been evaluated over a number of selected pages extracted from the Salzinnes book. The evaluation requires manual creation of ground truth data since due to the specificity of the task there are no other pieces of software known to the authors adapted to the tracking and reconstruction of such type of tetragram data. In the selected pages, the staff lines were carefully marked, linking the missing parts following the staff-slope trend and observing the limits (start and end points) of the staff lines, which are considered to run from the left side to the right side of each page.

In order to perform the evaluation, it must be taken into account that the ground truth creation task is a very tedious, eye-straining, prone-to-error and time-consuming one that requires much attention in order to create the best ground truth possible at the working resolution. This fact needs to be mentioned because, as a consequence, the ground truth has been observed to commonly deviate from the column-wise center of the images' staff lines as it is shown in the sample excerpt in Figure 9. This will be taken into account in order to perform a fair evaluation.

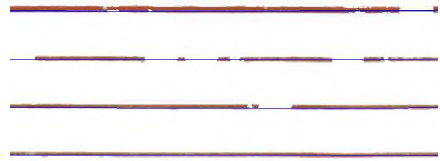

**Figure 9.** Example of staff line ground truth data drawn on the original image. Sample excerpt.

Data describing the ground truth created are shown in Table 1. In order to create the staff line ground truth, control points are set on the images obtained after the Pixelwise scheme with the aim that the spline curve passing through such points run through the vertical center of gravity of the corresponding staff line. The difference in the number of control points for different staff lines relies on the length of each staff line and the necessity of more or less control points so that the splines employed to characterize the staff lines actually lay on the real staff lines.

The control points are employed to create a set of ground truth staff lines by using spline curves running through them, the staff lines created are formed by a total number of 664,425 pixels. The locations of these pixels are employed to perform the evaluation and obtain the evaluation scores that will be described.

The evaluation is done by measuring the scores described in Table 2. The meaning of the scores is as follows:

- Detection of staff line: at a certain pixel location, the scheme presented marked the pixel as corresponding to a staff line, the binary image presented a level corresponding to the image foreground and a pixel of a ground truth staff line is found at a close vertical position (this idea will be defined later). This indicates a valid staff line detection/tracking at the pixel location.
- Interpolation of staff line: at a certain pixel location, the scheme presented marked the pixel as corresponding to a staff line, the binary image presented a level corresponding to the image background and a pixel of a ground truth staff line is found at a close vertical position. This indicates a valid staff line interpolation at the pixel location.
- Missed detection: at a certain location, a pixel of a ground truth staff line is found, the binary image presented a level corresponding to the image foreground and the scheme presented did not mark a pixel as corresponding to a staff line at any close vertical position. This indicates that a detection is missed.
- Missed interpolation: at a certain location, a pixel of a ground truth staff line is found, the binary image presented a level corresponding to the image background and the scheme presented did not mark a pixel as corresponding to a staff line at any close vertical position. In this case, the staff line needs to be interpolated/reconstructed, but the scheme proposed did not perform the interpolation properly or did not perform the interpolation at all.
- False detection: at a certain location, the scheme presented marked a pixel as corresponding to a staff line and the binary image presented a level corresponding to the image foreground but a pixel of a ground truth staff line was not found at a close vertical position.
- False interpolation: at a certain location, the scheme presented marked a pixel as corresponding to a staff line, the binary image presented a level corresponding to the image background at the same location and no ground truth staff line pixel was found at a close vertical position.

Note that additionally, we will use the term 'correct reconstruction' to refer to the union of the sets with correctly detected and correctly interpolated staff pixels.

**Table 1.** Summary of the ground truth data creation parameters.

| | |
|---|---|
| Staves | 69 |
| Staff lines | 276 |
| Total # control points | 3919 |
| Mean control points per staff line | 14.2 |
| Standard deviation of # control points per staff line | 3.31 |
| Minimum # control points per staff line | 6 |
| Maximum # control points per staff line | 21 |
| Mean control points per staff | 56.8 |
| Standard deviation of # control points per staff line | 12.8 |
| Minimum # control points per staff line | 24 |
| Maximum # control points per staff line | 83 |

Due to the way in which they are created, the ground truth staff lines may not lay at the vertical center of the images' staff lines. Consequently, in order to perform the evaluation, this fact is taken into account by considering correct detection and interpolation when the ground truth pixel is at a position that is close to the reconstructed pixel in the vertical direction, namely, the maximum vertical separation allowed for that estimated average staff line width at the current image.

The behavior of such vertical distance between the ground truth pixels and the detected pixels was observed in order to gather some detailed information about it. The histogram found is shown in Figure 10. This figure shows that the maximum separation allowed between the ground truth and the reconstructed staves is enough to embrace all the situations that can happen in our scheme (note that the estimated thickness of the staff lines ranges between 6.50 and 12.86 px, with 10.25 as global average) and that the ground truth is mostly correctly created with its staff lines running through the vertical center of the (reconstructed) staff lines or displaced only 1 pixel in most cases. Regarding the separation between staff lines, the estimated average separation between them in a staff ranges between 51.33 and 62.506 px, with 62.04 on average.

**Table 2.** Description of the staff line detection and tracking scheme performance score. At a certain pixel location, value '1' indicates that: (a) Detection: a staff line detected reconstructed goes over such location; (b) Ground truth: a ground truth staff line goes over such location; (c) Image: a foreground pixel in the binary image is found at such location.

| Interpretation | Detection | Ground Truth | Image |
|---|---|---|---|
| Detection of staff line | 1 | 1 | 1 |
| Interpolation of staff line | 1 | 1 | 0 |
| Missed detection | 0 | 1 | 1 |
| Missed interpolation | 0 | 1 | 0 |
| False detection | 1 | 0 | 1 |
| False interpolation | 1 | 0 | 0 |

The result of the evaluation of the performance of the scheme developed using the described ground truth is shown in Table 3. Observe that the percentage of staff line pixels (all the pixels in the reconstructed staff lines) is very close to 100%: 98.10%, this difference is partly due to the slightly smaller length of the reconstructed staff lines that the scheme produces. Note that this number could also be larger than 100% if the number of reconstructed staff line pixels becomes larger than the number of ground truth pixels.

The percentage of correctly detected pixels, which includes correctly detected and interpolated pixels, is also close to 100%: 97.55%. This fact points that the system performs correctly. Observe that the percentage of interpolated pixels is small, this is not an indication of wrong behavior but of the fact that interpolation occurs in a small percentage of the cases.

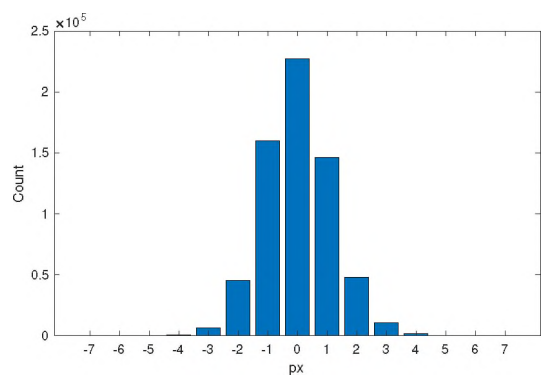

**Figure 10.** Histogram of the vertical separation between ground truth and detected pixels.

The error measures were low for all the types considered, with the largest one corresponding to the case of missed detections (2.17%), this can be related to the tight length of the reconstructed staff lines, which was previously mentioned.

The specificity of the task considered does not allow to find other works aimed at the same objective, however, in order to obtain some data for comparison, we have employed the schemes by Carter and Bacon [30] and Cardoso et al. [31] implemented by S. Vinitsky (http://pages.cs.wisc.edu/~vinitskys/omr/), as well as his machine-learning-based scheme for staff line removal. These schemes are built for staff line removal on music scores, so, we use as input for these schemes images drawn by the Pixelwise scheme containing both staves and musical symbols so that the schemes perform correctly. The combination of images shown in Figure 1b,d serves as an illustration of such images. The images are binarized by using Otsu's method [23] before the start of the staff search procedure.

The evaluation of the run-length scheme [30] draws the following numbers: correctly reconstructed pixels: 41.09%, correctly detected pixels: 40.73%, correctly interpolated pixels: 0.36%. The algorithm finds staff lines but it is not designed to create full staff lines based on interpolation, detection and tracking. An example of the results achieved by this scheme is shown in Figure 11a. The shortest path scheme attains 2.35% of correctly reconstructed pixels on the evaluation dataset. The scheme fails to detect staves in our particular scenario, this is the reason for such low number (no image is shown in Figure 11 since no staff lines were detected). The clustering algorithm by S. Vinitsky based on the utilization of feature vectors defined upon $9 \times 9$ tiles centered at each image pixel and the usage of the k-mean algorithm with Euclidean distance attains 91.85% of pixels correctly reconstructed, with 86.66% and 5.19% of pixels correctly detected and interpolated, respectively. This scheme successfully tracks most of the staff lines but gaps are not interpolated and the tracked line deviates from the actual staff line due to the presence of musical objects. As a matter of fact, a number of pixels are marked as correctly detected although they separate from the ground truth a notably observable distance, see Figure 11b. The reason for this behavior relies on the allowed detection displacement established in the evaluation scheme. Also, note that this scheme detects 107.29% of pixels with respect to the ground truth, which means that it marks as staff lines more pixels than in the ground truth. Figure 11c shows the result obtained by the proposed scheme on the top three staves on folio 7r of Salzinnes book.

**Table 3.** Evaluation results.

|  | Number | % w.r.t the Number of Ground Truth Pixels |
| --- | --- | --- |
| staff line pixels | 651,840 | 98.10% |
| correctly reconstructed pixels | 648,133 | 97.55% |
| correctly detected pixels | 597,173 | 89.88% |
| correctly interpolated pixels | 50,960 | 7.67% |
| missed detections | 14,389 | 2.17% |
| missed interpolations | 1903 | 0.29% |
|  |  | **% w.r.t the Number of Staff Line Pixels** |
| false detections | 1678 | 0.26% |
| false interpolations | 2029 | 0.31% |

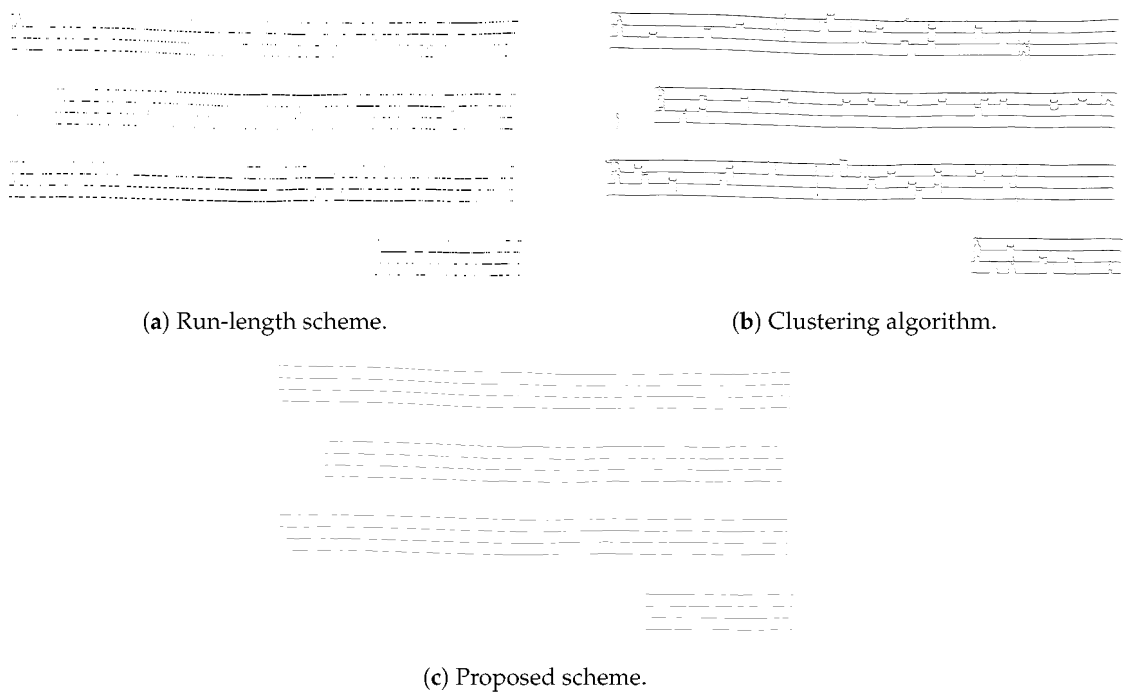

(a) Run-length scheme.　　　　　　　　　　　(b) Clustering algorithm.

(c) Proposed scheme.

**Figure 11.** Illustration of the behavior of selected schemes on folio 7r of Salzinnes book; detail of the top three staves.

## 5. Conclusions

In this paper, the problem of music staff tracking and reconstruction has been considered. In spite of the importance of staff lines for the correct interpretation of musical symbol on scores, OMR commonly consider their removal in the first stages to facilitate the identification of other musical objects. Making use of the output of such a system, we have described a processing scheme that is capable of successfully reconstructing the staff lines of handwritten scores from an ancient book.

The reconstructed staves are useful not only for the interpretation of the music symbols regarding their relative position with respect to the staff line, but also to improve the visualization of the manuscripts for end users, especially when the staff lines are broken in the original manuscript.

The processing scheme is based on the rational utilization of common image processing stages, a tracking scheme devised to make use simultaneously of the four tetragram lines of the staves under consideration and the selection of simple system parameters adapted to the task at hand. The evaluation performed shows remarkable performance but, still, there is room progress, like the definition of methods for the automatic selection of system parameters, which should make the proposed scheme immediately usable to perform the task considered on other datasets.

**Author Contributions:** Conceptualization: L.J.T., I.B., A.M.B. and I.F.; Data curation: L.J.T., I.B., A.M.B. and I.F.; Formal analysis: L.J.T., I.B., A.M.B. and I.F.; Funding acquisition: L.J.T., I.B. and I.F.; Investigation: L.J.T., I.B., A.M.B. and I.F.; Methodology: L.J.T., I.B., A.M.B. and I.F.; Project administration: I.B. and I.F.; Resources: L.J.T., I.B. and I.F.; Software: L.J.T., I.B., and A.M.B.; Supervision: L.J.T., I.B., A.M.B. and I.F.; Validation: L.J.T., I.B., A.M.B. and I.F.; Visualization: L.J.T., I.B., and A.M.B.; Writing original draft: L.J.T., I.B., A.M.B. and I.F.; Writing review and editing: L.J.T., I.B., A.M.B. and I.F. All authors have read and agreed to the published version of the manuscript.

**Funding:** Programa Operativo FEDER Andalucía 2014–2020: UMA18-FEDERJA-023; Universidad de Málaga, Campus de Excelencia Internacional Andalucía Tech.

**Acknowledgments:** This work has been funded by Programa Operativo FEDER Andalucía 2014–2020 under Project No. UMA18-FEDERJA-023 and Universidad de Málaga, Campus de Excelencia Internacional Andalucía Tech. This work has been partially carried out at McGill University, Montreal, Quebec, Canada.

**Conflicts of Interest:** The authors declare no conflict of interest.